# Unveiling Photoluminescence Signatures of Magneto-Optical Coupling in Layered Hybrid Manganese Chloride Perovskites

*Yaiza Asensio, Samuele Mattioni, Daniel Vaquero, Cédric A. Cordero-Silis, Houman Bahmani Jalali, Dorwal Marchelli, Marco Gobbi, Fèlix Casanova, Francesco Di Stasio, Marcos H.D. Guimarães, Luis E. Hueso,\* Beatriz Martín-García\**

Y. Asensio, S. Mattioni, F. Casanova, L.E. Hueso, B. Martín-García
CIC nanoGUNE BRTA, Tolosa Hiribidea, 76, 20018 Donostia-San Sebastián, Basque Country, Spain
Email: l.hueso@nanogune.eu, b.martingarcia@nanogune.eu

Y. Asensio, S. Mattioni
Departamento de Polímeros y Materiales Avanzados: Física, Química y Tecnología,
University of the Basque Country (EHU), Donostia-San Sebastian, 20018,
Spain

D. Vaquero, C.A. Cordero-Silis, M.H.D. Guimarães
Zernike Institute for Advanced Materials, University of Groningen, Groningen, Netherlands

H. Bahmani Jalali, D. Marchelli, F. Di Stasio
Photonic Nanomaterials, Istituto Italiano di Tecnologia, Via Morego 30, 16163 Genova, Italy

M. Gobbi
Materials Physics Center CSIC-/EHU, 20018 Donostia-San Sebastián, Spain

M. Gobbi, F. Casanova, L.E. Hueso, B. Martín-García
IKERBASQUE, Basque Foundation for Science, 48013 Bilbao, Basque Country, Spain

**Abstract text.**

**Abstract.** Understanding the interplay between magnetic ordering and light emission is crucial for developing magneto-optical technologies. However, this phenomenon is poorly understood since observations of this coupling vary significantly across materials. In this context, hybrid organic-inorganic metal halide perovskites (HOIPs) that incorporate $Mn^{2+}$ ions are a chemically and structurally tunable platform for exploring this phenomenon, since they exhibit magnetic ordering and photoluminescence (PL) emission. Here, we study two antiferromagnetic Mn-based HOIPs with different organic cations that result in distinct lattice stiffness, $Mn^{2+}$-$Mn^{2+}$ distance and octahedral distortion. Temperature-dependent PL excitation spectroscopy reveals changes in crystal field splitting energy and Racah parameters well above the Néel temperature ($T_N$), indicating the emergence of $Mn^{2+}$-$Mn^{2+}$ magnetic interactions prior to reach long-range magnetic ordering. These variations align with the observed changes in temperature-PL evolution. The compound with a more rigid lattice shows stronger changes closer to $T_N$, suggesting combined effects of magnetic polarons and spin-canting. In contrast, magnetic polaron-induced magnetic modifications prevail in the HOIP with a softer lattice. These results reveal the complexity of the magneto-optical coupling in Mn-based HOIPs and provide new insights into this field extensible to other 2D materials that exhibit this phenomenon with potential for advanced magneto-optical applications.

## 1. Introduction

Manganese (Mn)-based materials have emerged as promising candidates for next-generation optoelectronic and spintronic applications due to their interesting physical properties. Recent research has mainly focused on exploring magnetism in compounds such as layered manganese phosphorus trichalcogenides ($MnPS_3$, $MnPSe_3$, $MnPTe_3$),[1–3] manganese dichalcogenides ($MnS_2$, $MnSe_2$, $MnTe_2$,)[4–6] and Janus-type chalcohalides (MnSBr, MnSI, MnSeCl).[7] As other Mn-based systems, including halide perovskites and Mn-doped semiconductors,[8–12] some of them combine magnetic ordering with broadband photoluminescence (PL) emission. As a result, their potential for magneto-optical coupling, defined as the influence of magnetic ordering on PL emission, has become one of the most intriguing aspects of these materials. This coupling is particularly relevant for applications in spin-photonics,[13–15] magneto-optical memory,[16,17] thermal and pressure magneto-PL sensing,[18–20] and quantum information technologies,[21,22] where the coexistence of magnetic and optical responses enables simultaneous control of spin and light.

This multifunctionality of Mn-based materials arises from the properties of the $Mn^{2+}$ ions themselves. Their PL emission originates from spin-forbidden *d-d* transitions[23,24] and is highly sensitive to the strength and symmetry of the local crystal field. In particular, tetrahedral coordination typically results in green PL emission, whereas octahedral coordination is associated with orange to red PL.[25] Simultaneously, the half-filled $3d^5$ configuration of $Mn^{2+}$ (S = 5/2) gives rise to pronounced magnetic behavior, usually mediated by superexchange interactions following Goodenough–Kanamori rules.[26–28] Small lattice distortions such as changes in bond lengths or angular deviations can significantly modify the crystal field and magnetic exchange pathways, affecting both PL and magnetism.

However, despite this connection, observations of magneto-optical coupling remain inconsistent. Each material exhibits different PL signatures near its magnetic transition, ranging from shifts in emission intensity, peak position or linewidth, to the emergence of new bands. In some cases, such changes appear well above the magnetic transition temperature, suggesting that magneto-optical coupling may involve additional interactions beyond long-range magnetic order.[29–31] Similar variations have been observed in other systems, such as doped semiconductors and transition metal dichalcogenides (TMDs).[32–34] Notably, most magneto-optical studies have focused on Mn-doped semiconductors,[35–42] in which magnetic ions are sparsely distributed in a non-magnetic framework. This configuration results in a highly inhomogeneous magnetic environment, complicating the interpretation of PL responses. To advance this research, it is therefore necessary to explore structurally well-defined systems that enable controlled investigation of the coupling between magnetic order and optical emission.

In this regard, hybrid organic-inorganic metal halide perovskites (HOIPs) containing $Mn^{2+}$ ions offer a unique platform for studying these effects. These materials consist of inorganic layers of corner-sharing $[MnX_6]^{4-}$ octahedra which are separated by organic cations. Their chemical and structural versatility enables precise control over the lattice geometry and magnetic exchange pathways through the choice of the metal cation, halide and the organic spacer.[43–46] In particular, the organic spacer plays a crucial role in modulating Mn–Mn distances, octahedral distortion, and the rigidity of the framework—factors that directly affect both the magnetic interactions and the local crystal field experienced by $Mn^{2+}$ ions. While the PL of Mn-based HOIPs has been extensively studied in terms of coordination geometry, its connection to magnetic ordering remains poorly understood. A few studies have already reported PL changes near magnetic transitions in these materials.[47–49] However, these studies used a single material[48,49] or compounds with very different crystal structures and PL behavior[47], making it difficult to compare them. Therefore, the origin of these features is still under debate.

In this work, we face these challenges by taking advantage of the structural flexibility of Mn-based HOIPs to investigate how magnetism influences PL emission. We focus on two layered compounds that differ only in their organic spacers: phenylethylammonium (PEA), a rigid aromatic cation, and ethylammonium (EA), a more flexible aliphatic counterpart. This variation modulates the stiffness of the crystal lattice, influencing both the local crystal field and the PL behavior.[50,51] Structural differences were evaluated using reported crystallographic data at various temperatures, revealing different octahedral distortion and Mn–Mn distances.[52–55] Magnetic measurements confirmed the antiferromagnetic nature of both compounds, with $T_N$ close to 45 K consistent with previously reported values.[43,56–58]

To evaluate the interplay between magnetic and optical properties, we employed temperature-dependent PL and photoluminescence excitation (PLE) spectroscopies.[39] While PL reveals the emission properties of the system, PLE directly probes its electronic structure through the crystal field splitting energy ($\Delta$) and Racah parameters (B and C).[59–62] These parameters are directly connected to orbital configuration and lattice distortions, both of which influence magnetic behavior.[63] Their evolution with temperature revealed a transition in the 60-80 K range, indicating magnetic interactions at temperatures above the $T_N$. Moreover, the two compounds showed different responses. The stiffer framework of $PEA_2MnCl_4$, combined with its larger spin canting, led to higher-magnitude PL changes closer to its $T_N$, suggesting a competition between the FM contributions and its intrinsic AFM order. In contrast, $EA_2MnCl_4$, with its softer and more flexible lattice, exhibited smaller PL deviations, with a maximum emerging well above $T_N$. These findings advance our understanding of spin-light coupling in Mn-based HOIPs and other 2D materials, opening new pathways for the development of next-generation magneto-optical technologies.

## 2. Results and discussion

$PEA_2MnCl_4$ and $EA_2MnCl_4$ HOIP bulk crystals were synthesized by mixing $MnCl_2$ with the corresponding monoammonium cations, PEA and EA, in polar solvents (see Experimental section).[43] Photographs of the as-synthesized materials are shown in **Figure S1**. Additionally, we confirm the formation of the compounds and their layered structures by X-ray diffraction (XRD) (**Figure S2**). We select these two HOIPs, $PEA_2MnCl_4$ and $EA_2MnCl_4$, based on the distinct steric characteristics of their organic cations: PEA, a bulkier aromatic group, and EA, a smaller and more flexible aliphatic one. Both cations have a terminal $NH_3^+$ group, which interacts with the inorganic framework via N-H···Cl hydrogen bonds.[64] These N-H···Cl bonds influence the structural parameters of the $[MnCl_6]^{4-}$ octahedra (**Table 1**).[52–55] Indeed, PEA incorporates an additional phenyl ring connected to the ethyl chain, resulting in a significantly bulkier cation that also interact with neighboring cations via π-π stacking. These non-covalent interactions lead to greater rigidity in the organic layers and, in turn, increase in the stiffness of the lattice. We assess the lattice rigidity by evaluating the atomic displacement parameters in both compounds[52–55] (see Table S1). Atomic displacements are distinctly larger for $EA_2MnCl_4$ than for $PEA_2MnCl_4$ in both the organic and inorganic sublattices, confirming a stiffer lattice in the $PEA_2MnCl_4$ HOIP.[51,65] Therefore, these differences between the organic cations are likely to influence not only the crystal structure, but also the magnetic and optical behavior of the resulting materials.[43,50,51]

The crystallographic structures of both HOIPs are shown in **Figure 1**, revealing that interlayer distance of $PEA_2MnCl_4$ nearly doubles $EA_2MnCl_4$ (Table 1), consistent with the larger volume of the PEA cation. From the crystallographic data, we evaluate the lattice distortion induced by the different organic cations in terms of the mean octahedral quadratic elongation parameter ($\lambda_{oct}$) and the bond angle variance ($\sigma_{oct}$) (Table 1).[66,67] Despite the phase transition between two orthorhombic phases (*Cmce* to *Pbca*) near 230 K[55] for the $EA_2MnCl_4$ compound, the octahedral distortion is greater in this HOIP, as indicated by the larger $\lambda_{oct}$ values, than in $PEA_2MnCl_4$.

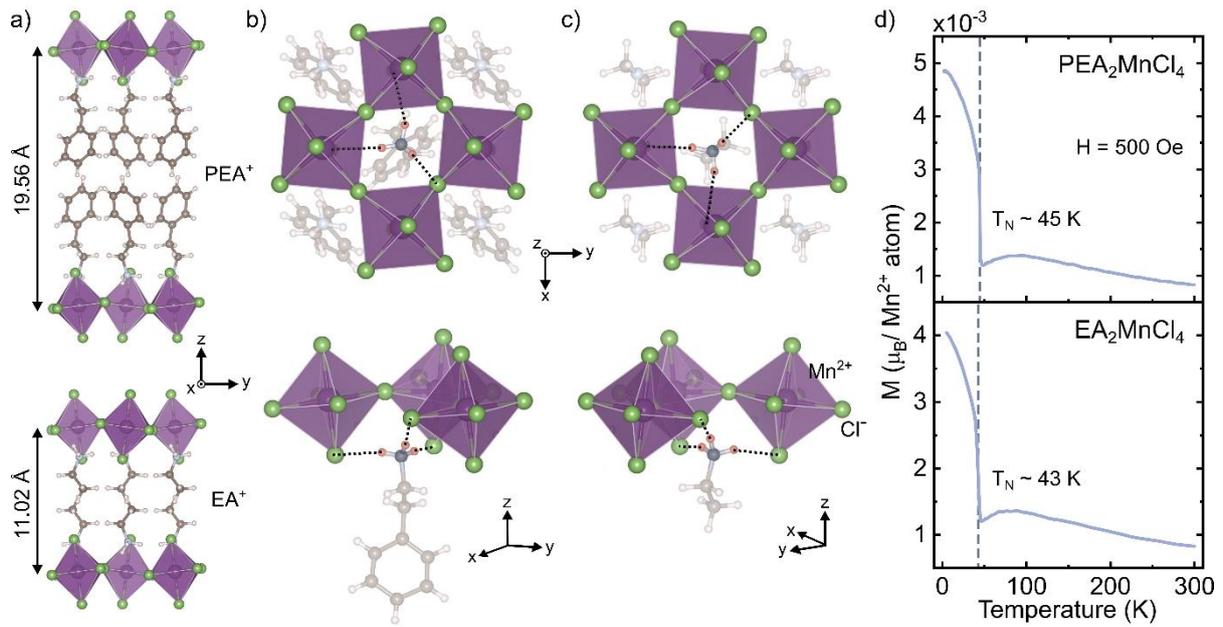

**Figure 1.** a-c) Schemes of the crystal structure of PEA$_2$MnCl$_4$ and EA$_2$MnCl$_4$, drawn using VESTA software with available crystallographic data at 100 K and 126 K, respectively.[52,55] b) top and side view of the hydrogen bonding between the NH$_3^+$ groups and the Cl$^-$ anions in PEA$_2$MnCl$_4$. c) shows equivalent views for EA$_2$MnCl$_4$. d) Temperature-dependent magnetization curves measured at 500 Oe for both compounds.

**Table 1.** Bond distances and angles between Mn$^{2+}$ and Cl$^-$ in PEA$_2$MnCl$_4$ and EA$_2$MnCl$_4$, extracted from reported crystal structure[52–55] and using VESTA software[67].

| Material | T [K] | Space group | Interlayer Mn-Mn distance [Å] | Intralayer Mn-Mn distance [Å] | Intralayer Mn-Cl-Mn angle [°] | Intralayer Cl-Mn-Cl angle [°] | $\lambda_{oct}$ | $\sigma_{oct}$ [°$^2$] | CCDC |
|---|---|---|---|---|---|---|---|---|---|
| PEA$_2$MnCl$_4$ | 300 | Orthorhombic (*Pbca*) | 19.70 | 5.12 | 168.50 | 90.23 | 1.0008 | 0.69 | 2411272 |
|  | 100 | Orthorhombic (*Pbca*) | 19.56 | 5.08 | 164.92 | 90.34 | 1.0005 | 0.73 | 1583818 |
| EA$_2$MnCl$_4$ | 300 | Orthorhombic (*Cmce*) | 11.04 | 5.16 | 171.58 | 91.05 | 1.0012 | 0.43 | 1148001 |
|  | 126 | Orthorhombic (*Pbca*) | 11.02 | 5.12 | 165.994 | 91.88 | 1.0013 | 2.38 | 1148005 |

Specifically, their Cl-Mn-Cl bond angles deviate by ~1° from the ideal 90°, compared to only ~0.3° in PEA$_2$MnCl$_4$. According to previous studies,[68,69] a Cl–Mn–Cl angles closer to 90° favor stronger ferromagnetic (FM) exchange interactions in antiferromagnetic (AFM) materials. In addition, intralayer Mn–Mn distance is shorter in PEA$_2$MnCl$_4$, which further enhances magnetic exchange. Therefore, these structural parameters suggest a stronger in-plane magnetic exchange in the PEA-based compound, in agreement with literature reports describing both materials as AFM with spin canting, but with a larger canting angle in PEA$_2$MnCl$_4$.[43,56,58] Both compounds also exhibit similar Néel temperature values, $T_N$ of ~45 K for PEA$_2$MnCl$_4$ and ~43 K for EA$_2$MnCl$_4$ determined from the temperature-dependent magnetization curves (Figure 1d), and in agreement with values reported in literature.

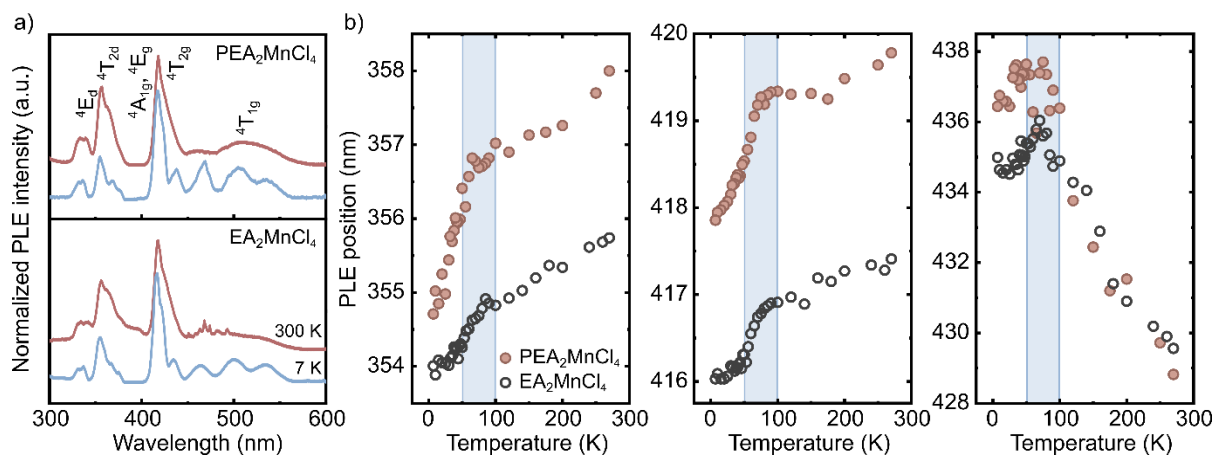

**Figure 2**: a) Photoluminescence excitation spectra for PEA$_2$MnCl$_4$ and EA$_2$MnCl$_4$ at 300 K and 7 K and their b) temperature-dependent evolution of $^4E_d$, $^4A_{1g}$-$^4E_g$ and $^4T_{2g}$ transitions in the range of 7-270 K.

While magnetic measurements confirm AFM ordering in both compounds, they do not directly reveal how these magnetic interactions relate to the electronic states of Mn$^{2+}$. To gain insight into them and their temperature-dependent behavior, we performed PLE measurements above and below the $T_N$ (**Figure S3**). For both 2D HOIPs under study, the shape of PLE spectra is consistent with that reported for other octahedrally coordinated Mn$^{2+}$ compounds (S=5/2, $d^5$ configuration – high spin).[70–72] At room temperature, in PEA$_2$MnCl$_4$ we observe five excitation bands centered at ~358, ~364, ~420, ~429 and ~516 nm, which correspond to transitions from the $^6A_{1s}$ ground state to $^4E_d$, $^4T_{2d}$, $^4A_{1g}$-$^4E_g$, $^4T_{2g}$ and $^4T_{1g}$ excited states, respectively (**Figure 2**.a).[12,70–72] In the case of EA$_2$MnCl$_4$, we observe the same transitions but slightly blueshifted (Figure 2b). We were able to properly monitor $^4E_d$, $^4A_{1g}$-$^4E_g$ and $^4T_{2g}$ transitions in the whole temperature range studied (Figure 2b and SI Figure S3).

For both PEA$_2$MnCl$_4$ and EA$_2$MnCl$_4$, the $^4E_d$ and $^4A_{1g}$-$^4E_g$ transitions exhibit a progressive blueshift as the temperature decreases. Notably, this shift becomes more pronounced below 60-80 K. In contrast, the $^4T_{2g}$ transition shows a redshift at higher temperatures, followed by a blueshift below ~80 K. Although this temperature range does not coincide with the $T_N$ of either compound, it may be attributed to a gradual magnetic transition as reported in other AFM materials such as MnS through PLE characterization,[39] and CrSBr and MnPS$_3$ using PL spectroscopy.[32–35]

Interestingly, a comparison of the results for the two compounds reveals that PEA$_2$MnCl$_4$ exhibits a larger overall temperature-dependent shift than EA$_2$MnCl$_4$ for all monitored excited states. This difference in overall temperature-dependent shift likely originates from the increased lattice rigidity imposed by PEA cations.[51,65] Nonetheless, in both materials, the extent of the shift also varies among transitions, which could be related to their crystal field dependence. Indeed, decreasing temperature enhances the crystal field strength due to lattice contraction (see reduction of Mn-Mn in-plane and interlayer distances in Table 1).[61,62,73–75] This stronger crystal field interacts with the Mn$^{2+}$ $d$-orbitals, leading to degeneracy and modifying the energy levels through what is defined as the crystal field splitting energy ($\Delta$). The evolution of these energy levels as a function of $\Delta$, relative to the Racah parameter B (which accounts for electron–electron repulsion),[62,63] is well described by the Tanabe–Sugano diagram (**Figure S4**).[76,77] The transitions that present such a strong dependence on $\Delta$ are $^4T_{1g}$, $^4T_{2g}$ and $^4T_{2d}$, while $^4A_{1g}$-$^4E_g$ and $^4E_d$ remain unaffected. Certainly, we observe the largest overall temperature-dependent shift in the $^4T_{2g}$ transition, consistent with its high $\Delta$ dependence. On the other hand, $^4E_d$ and $^4A_{1g}$-$^4E_g$ transitions, despite of not being sensitive to $\Delta$, still exhibit small shifts in temperature, likely due to variations in the Racah parameters.[19,20]

To investigate this scenario and gain insight into the relationship between the observed PLE transitions and changes in both interelectronic repulsion and crystal field effects, we calculated the Racah parameters B and C, as well as $\Delta$, from the temperature-dependent PLE spectra. This analysis was performed using the matrix solutions derived from the Tanabe-Sugano model for $d^5$ configuration,[76,77] incorporating the α correction parameter developed by Trees (more details in S.I.). [61,78–82] **Figure 3**.a-c shows the temperature evolution of B, C/B and $\Delta$ parameters for both compounds, with values in close agreement with the expected ones for Mn$^{2+}$-based compounds:[61] B=570-790 cm$^{-1}$, C=3200-3770 cm$^{-1}$, C/B=4-6.5. As representative example, Figure 3d displays the specific Tanabe-Sugano diagram for PEA$_2$MnCl$_4$ at 7 K, based on the PLE spectrum. Using the parameters obtained (B=756.1 cm$^{-1}$, C=2977.1 cm$^{-1}$ and $\Delta$=8432.9 cm$^{-1}$), we estimated $\Delta$/B=11.2 (dotted line), which provides a direct comparison between the calculated energy levels and the measured PLE transitions.

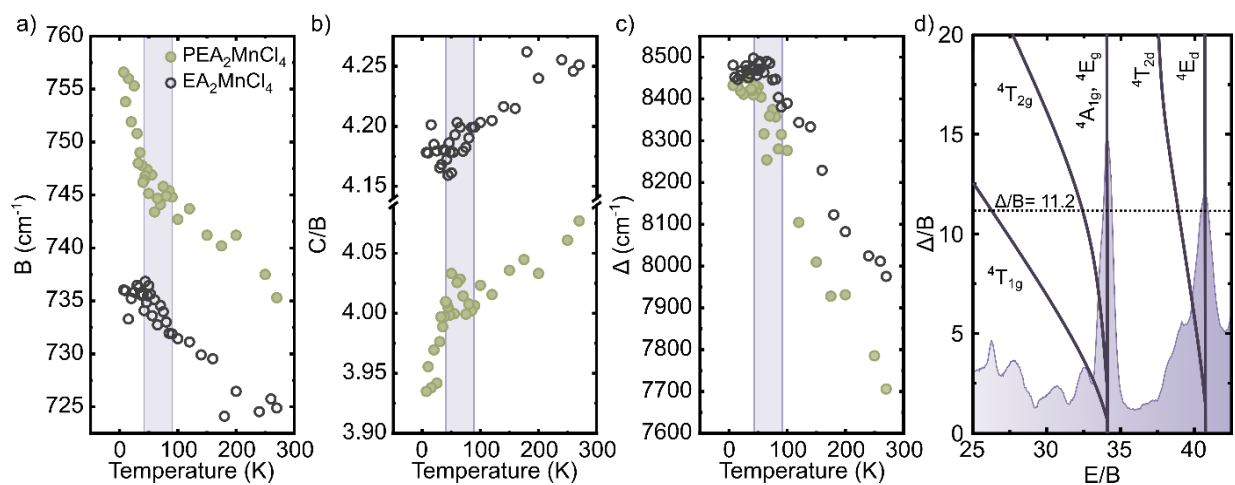

**Figure 3**: Temperature-dependent evolution of Racah parameters a) B and b) C/B related to electrostatic interactions between the electrons occupying the *d*-orbitals, and c) crystal field energy splitting Δ. The regions shaded in purple indicate the temperature at which changes in trend are observed. d) Tanabe-Sugano diagram for $PEA_2MnCl_4$ based on its PLE spectrum at 7 K. Only the energy levels associated with observed transitions are shown for clarity.

The Racah parameters B and C, [59,60,75] quantify the electrostatic interactions between the electrons occupying the *d*-orbitals, with B directly related to the interelectron repulsion.[62,63] As shown in Figure 3a-b, both compounds exhibit a systematic increase in B when cooling, indicating a stronger repulsion between the electrons at lower temperatures. $PEA_2MnCl_4$ presents higher B values compared to $EA_2MnCl_4$, suggesting a stronger electronic localization. Notably, a change in slope is observed around 50–70 K, below which B increases more rapidly. This change is especially more marked in $PEA_2MnCl_4$—with a shift of ~12 cm$^{-1}$, compared to ~5 cm$^{-1}$ in $EA_2MnCl_4$—and indicates modifications in the orbital configuration, likely related to magnetic ordering. Therefore, these results suggest the presence of magnetic ordering in the materials above the $T_N$, which may not be detectable with magnetic measurements as the ones shown in Figure 1.d. The C parameter decreases when lowering the temperature for both materials, showing a change in its trend around the same temperature range as B parameter (**Figure S5**). This coordinated trend of B and C correlates with the observed temperature shifts in $^4E_d$ and $^4A_{1g}$-$^4E_g$ transitions responding to variations to Racah parameters. The C/B ratio, which determines the reference energy of the levels in the Tanabe–Sugano diagram, also decreases when cooling and presents lower values for $PEA_2MnCl_4$ than for $EA_2MnCl_4$. As previously observed in the temperature-dependent evolution of PLE transitions, a distinct change in trend is observed well above the $T_N$ for both compounds.

Moreover, the calculated values for $\Delta$ show how the orbital energy difference between $e_g$ and $t_{2g}$ levels evolves with temperature.[62] In both compounds, $\Delta$ increases from room temperature to approximately 60–80 K, consistent with lattice contraction effects. Under 60–80 K, $\Delta$ becomes nearly constant for both materials, suggesting that competing interactions begin to compensate for the crystal field enhancement caused by lattice contraction. Such interactions could arise from magnetically driven structural distortions or exchange interactions at higher temperatures than the $T_N$,[47,83] as supported by the simultaneous changes observed in the Racah parameters.

As previously discussed, the transition from $^6A_{1s}$ ground state to $^4T_{1g}$ excited state is strongly dependent on $\Delta$ (Figure 3.d). Its corresponding $^4T_{1g} \to {^6A_{1s}}$ emission can be directly probed via PL spectroscopy, typically appearing at lower energies due to vibrational relaxation and non-radiative losses. According to the Tanabe-Sugano model, a redshift in this PL emission should occur with decreasing temperature as a result of the increasing crystal field strength. However, given the complex evolution of both $\Delta$ and the Racah parameters observed from the PLE measurements, deviations from this behavior are possible.

**Figure 4** presents the temperature-dependent PL spectra (complete temperature range is shown in **Figure S6**) together with the evolution of the peak position and full width at half maximum (FWHM) for both compounds in the range 5 –300 K. Notably, the PL emission of $PEA_2MnCl_4$ is slightly redshifted (609.5 nm) compared to that of $EA_2MnCl_4$ (606.7 nm), reflecting the structural influence of the organic spacer on lattice rigidity, Mn-Mn distances and octahedral distortion (Table 1), as previously observed also in the corresponding PLE transitions.

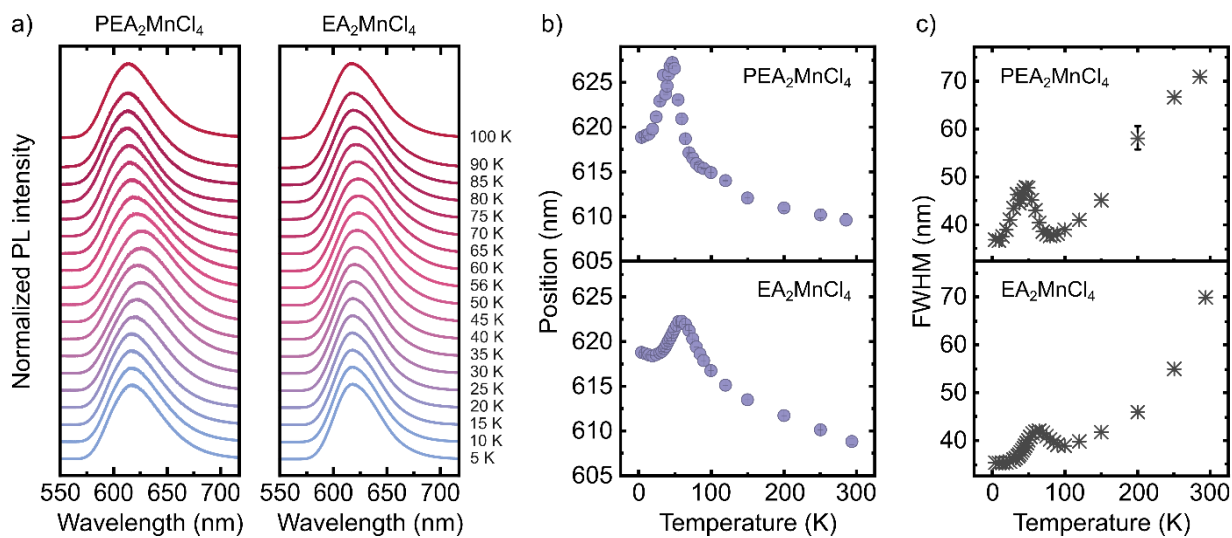

**Figure 4**: a) PL emission spectra for PEA$_2$MnCl$_4$ and EA$_2$MnCl$_4$ in the temperature range of 5-100 K. b) PL peak position and c) FWHM as a function of the temperature (5-300 K) for both compounds.

At higher temperatures, both materials display a redshift of the peak position due to the lattice contraction and increasing crystal field strength, and a narrowing of the FWHM due to reduced phonon activity.[43,74,84] However, below approximately 60-80 K—the temperature range where we identified deviations in Racah parameters and crystal field splitting energy by PLE characterization —for both PEA$_2$MnCl$_4$ and EA$_2$MnCl$_4$ we observe a peak in the temperature-dependent PL evolution. As PLE results pointed out, these changes under this specific temperature range could be ascribed to the magnetic interactions between neighboring Mn$^{2+}$ ions before reaching the T$_N$ (~45 K and 43 K for PEA$_2$MnCl$_4$ and EA$_2$MnCl$_4$, respectively).[39,47,83] One possible explanation involves the formation of magnetic polarons,[29–31,47] quasiparticles formed when photoexcited carriers locally align the spins of the magnetic ions, creating a small region of local magnetic ordering, typically ferromagnetic (FM),[29–31,47] that becomes spectroscopically active at temperatures typically below ~100 K. This photoinduced magnetic behavior has been widely reported in diluted magnetic semiconductors[14,30,31,85–87] where the formation of band-to-band excitons locally aligns surrounding Mn$^{2+}$ spins via interactions with longitudinal optical phonons. Although the electronic structure of Mn$^{2+}$ HOIPs does not favor the formation of delocalized excitons as in conventional semiconductors, analogous photoinduced effects can still occur through localized charge-transfer excitations, leading to charge-transfer magnetic polarons. In these processes, an electron is promoted from a halide *p* orbital to a Mn 3*d* orbital, forming a ligand-to-metal charge-transfer state. This introduces a Mn center with a different electronic configuration within the Mn$^{2+}$ lattice, locally modifying exchange interactions and potentially aligning nearby spins. [29,47,88]

Notably, magnetic polaron formation can occur even above the $T_N$, inducing local alignment of the random-oriented spins of the paramagnetic phase. Once the system enters in the AFM phase, this AFM alignment can be modified locally by the excitations, either enhancing or suppressing it. The emergence of magnetic polarons would explain why these magnetic interactions are not detected simply by M(T) or M(H) measurements, as they are induced by optical excitations.

In detail, deviations from the high-temperature PL behavior appear below ~80 K in both compounds. The PL emission redshifts more rapidly, while the FWHM reverses its trend and begins to broaden. These changes may reflect the formation of magnetic polarons inducing magnetic interactions.[47] The maximum of these peak-shape behaviors appears around 48 K for PEA$_2$MnCl$_4$ and 59 K for EA$_2$MnCl$_4$, below which the PL emission suddenly blueshifts and the FWHM narrows, which could be consistent with the emergence of AFM ordering in the materials. Although magnetic measurements report a $T_N$ (45 K and 43 K) slightly lower than these temperatures where PL maximum change occurs, similar differences have been reported in other Mn-based systems,[32,33,35] where optically detected magnetic transitions appear more gradually and at slightly elevated temperatures. Additionally, a subtle change in the slope is observed in both peak position and FWHM near 10 K for PEA$_2$MnCl$_4$ and 22 K for EA$_2$MnCl$_4$, possibly indicating that the AFM ordering is completely established or that the EMP saturates. Most interestingly, the maximum of both PL peak position and FWHM for PEA$_2$MnCl$_4$ are larger in magnitude compared to those of EA$_2$MnCl$_4$, consistent with the larger temperature-dependent shifts observed in the PLE transitions of this compound.

The observed differences between the two HOIPs regarding the different temperatures for the maximum and the change at the lowest temperature may be attributed to the larger spin canting angle in PEA$_2$MnCl$_4$. This phenomenon introduces weak ferromagnetism in the material, which may compete next to the magnetic interactions induced by the magnetic polarons against the intrinsic AFM interactions. As a result, PEA$_2$MnCl$_4$ shows the AFM effect on its PL much later (maxima of the peaks), as well as the complete AFM order. Nevertheless, it is worth noting that the optically induced magnetic polarons seem to importantly modify the magnetic behavior of both HOIPs.

## 3. Conclusion

In this work, we investigated the interplay between crystal structure, magnetism and optical properties in two layered Mn-based HOIPs, PEA$_2$MnCl$_4$ and EA$_2$MnCl$_4$, with different organic spacers leading to a different lattice rigidity. While PEA imposes greater lattice stiffness, EA leads to a softer framework, resulting in different Mn-Mn distances and octahedral distortion degree for each compound. We confirmed the AFM nature of both compounds by magnetic measurements, exhibiting similar T$_N$ values (45 K and 43 K). Temperature-dependent PLE measurements showed changes in both crystal field splitting energy and Racah parameters well above T$_N$, indicating that magnetic driven interactions modified the Mn$^{2+}$ orbital environment before long-range magnetic ordering is reached. These electronic changes correlate well with changes in the temperature-dependent PL emission evolution in terms of peak position and linewidth, where we observe a maximum. In the case of PEA$_2$MnCl$_4$, with a more rigid lattice, stronger crystal field sensitivity, and larger spin canting, it shows more marked changes (larger magnitude of change) in temperature-dependent PLE and PL characteristics, with the maximum emerging at lower temperatures than EA$_2$MnCl$_4$, likely due to the combined magnetic effects of the magnetic polarons with the spin canting competing with AFM exchange. In EA$_2$MnCl$_4$, with a softer lattice, lower spin canting, and weaker crystal field sensitivity, the PL changes are less marked, with the maximum appearing well above the T$_N$, suggesting that optically induced magnetic polarons significantly modify its magnetic behavior. Therefore, this work demonstrates that the choice of organic spacer can effectively influence the magnetic and optical properties of HOIPs, while also revealing the intricate nature of the magneto-optical coupling in transition metal compounds. Beyond advancing the understanding of magneto-optical coupling in HOIPs, this study offers a versatile basis for exploring related phenomena in other 2D materials combining PL and PLE spectroscopies to detect magnetic effects towards the design of next-generation magneto-optical systems.

## 4. Experimental Section

*Synthesis of Mn-based HOIPs*: Bulk crystals, PEA$_2$MnCl$_4$ and EA$_2$MnCl$_4$, were synthesized using stoichiometric amounts of the precursors 2:1 molar ratio of organic molecule and MnCl$_2$ (63 mg, 99.99%, anhydrous beads 10 mesh, Sigma Aldrich-Merck) in polar solvents.[56] In the case of PEA$_2$MnCl$_4$, we mixed the amine, phenylethylamine (126 μL, purified by redistillation, ≥99.5%, Sigma Aldrich-Merck), with HCl (650 μL, 37%, Sigma Aldrich-Merck) to form the salt and the medium used was acetone:ethanol (2:1 mL). For $EA_2MnCl_4$ crystals, we used the organic salt EACl (81.5 mg, Sigma Aldrich-Merck) and 1 mL of ethanol as medium. In both cases, we stirred the solutions at 120 °C up to achieve a clear almost transparent solution. Then, the vial was immediately transferred to another hot plate set at 35 °C to allow the growth of the pale pink crystals (1-2 days).

*Photoluminescence spectroscopy characterization:* PL measurements were performed in a home-built micro-PL setup. A supercontinuum white laser (NKT Photonics SuperK EXTREME) was used as an excitation source and selecting the 532 nm wavelength (1 nm FWHM) using an AOTF (NKT Photonics SELECT). The laser power was kept below 1 mW for all measurements. The laser beam was first collimated and linearly polarized, and focused on the crystals using a 10x microscope long working distance objective to obtain a diffraction-limited spot of ~1 μm. The samples were mounted inside a Janis ST-500 flow cryostat under high vacuum (~5 x $10^{-7}$ mbar), with temperature control ranging from 5 K to 300 K. PL emission was collected in a backscattering geometry through the same objective, filtered with a 550 nm long-pass filter to remove residual excitation light, and finally analyzed using an Andor Shamrock 500i spectrometer, using an iDus 420 thermoelectrically cooled CCD detector with a 300 l/mm grating.

*Photoluminescence excitation spectroscopy characterization:* The PLE spectra of the bulk materials were collected via an Edinburgh FLS900 fluorescence spectrometer equipped with a Xe lamp and a monochromator for steady-state PL excitation, collecting at the peak emission. For the temperature-dependent PLE measurements, the samples were mounted in a closed-cycle cryostat (ARS) and the temperature was increased from 7 K to 270 K. The measurements at 300 K were performed outside the cryostat.

*Magnetic Characterization:* Magnetic measurements along the in-plane crystallographic direction were carried out on a Physical Properties Measurement System (PPMS) operating in the vibrating sample magnetometer mode. Magnetization as a function of temperature was measured between 5 and 300 K under constant magnetic field of 500 Oe. To ensure data reliability, at least three crystals of each material were measured. The magnetic transition temperature was determined using the first derivative of the M(T) curves, with an error of ±2 K.

*X-Ray Diffraction characterization*: XRD measurements were performed using an PANalytical Empyrean diffractometer equiped with a copper anode, operating with Cu K$\alpha_1$ (1.5406 Å) and Kα$_2$ (1.5443 Å) wavelengths to maximize the intensity of the diffracted beam. Reproducibility was confirmed by measuring at least three crystals of each compound, including samples from different batches.

**Acknowledgements**

This work is supported under Projects PID2021-122511OB-I00, PID2021-128004NB-C21, PID2024-157558NB-C21, PID2024-157558NB-C22 and PID2024-155708OB-I00 and under the María de Maeztu Units of Excellence Programme (Grant CEX2020-001038-M) funded by Spanish MICIU/AEI/10.13039/501100011033 and by ERDF/EU. Additionally, this work was carried out with support from the Basque Science Foundation for Science (IKERBASQUE), concretely, B.M.G. thanks IKERBASQUE HYMNOS project. Y.A. thanks the funding from Spanish MICIU/AEI/10.13039/501100011033 and ESF+ (PhD grant PRE2021-099999). S.M acknowledges support from the Department of Education of the Basque Government under the Pre-doctoral Programme for the Training of Non-doctoral Research Staff. B.M.-G. and M.G. thanks support from "Ramón y Cajal" Programme by the Spanish MICIU/AEI/10.13039/501100011033 and European Union NextGenerationEU/PRTR (grant nos. RYC2021-034836-I and RYC2021-031705-I, respectively). C.A.C.S. D.V. and M.H.D.G acknowledge the financial support from the European Union (ERC, 2D-OPTOSPIN, 101076932), the Zernike Institute for Advanced Materials, and the research program "Materials for the Quantum Age" (QuMat – registration number 024.005.006) which is part of the Gravitation program financed by the Dutch Ministry of Education, Culture and Science (OCW). Authors thank SGIker Medidas Magnéticas Gipuzkoa (UPV/EHU) for the technical and human support, concretely, to Dr. M. Ipatov.

**Data Availability Statement**

The data that support the findings of this study are available from the corresponding author upon reasonable request.

**References**

[1] N. Wang, J. Chen, Y. An, Q. Zhan, S.-J. Gong, *npj Spintronics* **2024**, *2*, 60.
[2] W. Zhou, G. Zheng, A. Li, D. Zhang, F. Ouyang, *Phys. Rev. B* **2023**, *107*, 035139.

[3]  D. Tezze, J. M. Pereira, D. Tutar, M. Ramos, J. Regner, P. Gargiani, F. Schiller, F. Casanova, A. Alegria, B. Martín-García, H. Sahin, Z. Sofer, M. Ormaza, L. E. Hueso, M. Gobbi, *Adv Funct Materials* **2025**, *35*, 2412771.
[4]  Y. Zhao, Y. Wan, C. Huang, P. Gu, X. Zhang, Y. Wu, M. Liu, Y. Li, K. Wang, E. Kan, *J. Am. Chem. Soc.* **2025**, jacs.5c10107.
[5]  L. Luo, Q. Sun, C. Jin, M. Li, R. Tan, Y. Dai, *J. Phys. Chem. Lett.* **2024**, *15*, 12181.
[6]  J.-W. Li, G. Su, B. Gu, *Phys. Rev. B* **2024**, *109*, 134436.
[7]  X.-F. Huang, K.-J. Li, Z. Wang, S.-B. Zhao, B. Shen, Z.-X. Chen, Y. Hou, *Applied Physics Letters* **2024**, *124*, 252402.
[8]  L. Chen, W. Yang, H. Fu, W. Liu, G. Shao, B. Tang, J. Zheng, *J Mater Sci* **2021**, *56*, 8048.
[9]  H. Xu, W. Liang, Z. Zhang, C. Cao, W. Yang, H. Zeng, Z. Lin, D. Zhao, G. Zou, *Advanced Materials* **2023**, *35*, 2300136.
[10] M. Mukhtar, B. S. Goud, Z. Ali, M. W. Shaid, B. Naz, Q. U. Ain, M. A. Assiri, S. Sonmezoglu, A. H. Rajpar, J. H. Kim, S. Aftab, *J. Alloys Compd.* **2025**, *1026*, 180365.
[11] R. Babu, I. López-Fernández, S. Prasanthkumar, L. Polavarapu, *ACS Appl. Mater. Interfaces* **2023**, *15*, 35206.
[12] Y. Asensio, H. Bahmani Jalali, S. Marras, M. Gobbi, F. Casanova, A. Mateo-Alonso, F. Di Stasio, I. Rivilla, L. E. Hueso, B. Martín-García, *Adv. Opt. Mater.* **2024**, 2400554.
[13] W. D. Rice, W. Liu, V. Pinchetti, D. R. Yakovlev, V. I. Klimov, S. A. Crooker, *Nano Lett.* **2017**, *17*, 3068.
[14] I. A. Akimov, T. Godde, K. V. Kavokin, D. R. Yakovlev, I. I. Reshina, I. V. Sedova, S. V. Sorokin, S. V. Ivanov, Yu. G. Kusrayev, M. Bayer, *Phys. Rev. B* **2017**, *95*, 155303.
[15] T. Neumann, S. Feldmann, P. Moser, A. Delhomme, J. Zerhoch, T. Van De Goor, S. Wang, M. Dyksik, T. Winkler, J. J. Finley, P. Plochocka, M. S. Brandt, C. Faugeras, A. V. Stier, F. Deschler, *Nat Commun* **2021**, *12*, 3489.
[16] G. Mackh, W. Ossau, D. R. Yakovlev, G. Landwehr, T. Wojtowicz, G. Karczewski, J. Kossut, *Acta Phys. Pol. A* **1995**, *88*, 849.
[17] V. F. Agekyan, P. O. Holz, G. Karczewski, V. N. Katz, E. S. Moskalenko, A. Yu. Serov, N. G. Filosofov, *Semiconductors* **2011**, *45*, 1301.
[18] E. Song, S. Ding, M. Wu, S. Ye, F. Xiao, S. Zhou, Q. Zhang, *Advanced Optical Materials* **2014**, *2*, 670.
[19] L. Nataf, F. Rodríguez, R. Valiente, J. González, *High Press. Res.* **2009**, *29*, 653.
[20] Y. Rodríguez-Lazcano, L. Nataf, F. Rodríguez, *Phys. Rev. B* **2009**, *80*, 085115.
[21] L. Besombes, Y. Léger, L. Maingault, D. Ferrand, H. Mariette, J. Cibert, *Phys. Rev. Lett.* **2004**, *93*, 207403.
[22] A. Kudelski, A. Lemaître, A. Miard, P. Voisin, T. C. M. Graham, R. J. Warburton, O. Krebs, *Phys. Rev. Lett.* **2007**, *99*, 247209.
[23] J. Orive, J. L. Mesa, R. Balda, J. Fernández, J. Rodríguez Fernández, T. Rojo, M. I. Arriortua, *Inorg. Chem.* **2011**, *50*, 12463.
[24] J. Ferguson, H. J. Guggenheim, Y. Tanabe, *J. Phys. Soc. Jpn.* **1966**, *21*, 692.
[25] Q. Zhou, L. Dolgov, A. M. Srivastava, L. Zhou, Z. Wang, J. Shi, M. D. Dramićanin, M. G. Brik, M. Wu, *J. Mater. Chem. C* **2018**, *6*, 2652.
[26] J. B. Goodenough, *J. Phys. Chem. Solids* **1958**, *6*, 287.
[27] J. B. Goodenough, *Phys. Rev.* **1955**, *100*, 564.
[28] J. Kanamori, *J. Phys. Chem. Solids* **1959**, *10*, 87.
[29] H. Peng, T. Huang, B. Zou, Y. Tian, X. Wang, Y. Guo, T. Dong, Z. Yu, C. Ding, F. Yang, J. Wang, *Nano Energy* **2021**, *87*, 106166.
[30] B. Zou, Y. Tian, L. Shi, R. Liu, Y. Zhang, H. Zhong, *J. Lumin.* **2022**, *252*, 119334.

[31] M. A. Kamran, B. Zou, K. Zhang, X. Yang, F. Ge, L. Shi, T. Alharbi, *Research* **2019**, *20019*.
[32] Y. Zhou, K. He, H. Hu, G. Ouyang, C. Zhu, W. Wang, S. Qin, Y. Tao, R. Chen, L. Zhang, R. Shi, C. Cheng, H. Wang, Y. Liu, Z. Liu, T. Wang, W. Huang, L. Wang, X. Chen, *Laser Photonics Rev.* **2022**, *16*, 2100431.
[33] Y. Xing, H. Chen, A. Zhang, Q. Hao, M. Cai, W. Chen, L. Li, D. Peng, A. Yi, M. Huang, X. Wang, J. Han, *Adv. Opt. Mater.* **2025**, *13*, 2402549.
[34] W. M. Linhart, M. Rybak, M. Birowska, P. Scharoch, K. Mosina, V. Mazanek, D. Kaczorowski, Z. Sofer, R. Kudrawiec, *J. Mater. Chem. C* **2023**, *11*, 8423.
[35] X. Yang, C. Pu, H. Qin, S. Liu, Z. Xu, X. Peng, *J. Am. Chem. Soc.* **2019**, *141*, 2288.
[36] E. Müller, W. Gebhardt, *Phys. Status Solidi B* **1986**, *137*, 259.
[37] M. M. Moriwaki, W. M. Becker, W. Gebhardt, R. R. Galazka, *Solid State Commun.* **1981**, *39*, 367.
[38] J. Ferguson, H. J. Guggenheim, Y. Tanabe, *Phys. Rev. Lett.* **1965**, *14*, 737.
[39] W. Heimbrodt, C. Bencke, O. Goede, H. -E. Gumlich, *Phys. Status Solidi B* **1989**, *154*, 405.
[40] B. Yang, X. Shen, H. Zhang, Y. Cui, J. Zhang, *Sci. Adv. Mater.* **2014**, *6*, 623.
[41] L. R. Bradshaw, J. W. May, J. L. Dempsey, X. Li, D. R. Gamelin, *Phys. Rev. B* **2014**, *89*, 115312.
[42] J. F. MacKay, W. M. Becker, J. Spaek, U. Debska, *Phys. Rev. B* **1990**, *42*, 1743.
[43] Y. Asensio, S. Marras, D. Spirito, M. Gobbi, M. Ipatov, F. Casanova, A. Mateo-Alonso, L. E. Hueso, B. Martín-García, *Adv. Funct. Mater.* **2022**, *32*, 2207988.
[44] Y. Asensio, L. Olano-Vegas, S. Mattioni, M. Gobbi, F. Casanova, L. E. Hueso, B. Martín-García, *Mater. Horiz.* **2025**, *12*, 2414.
[45] K. P. Lindquist, X. Xu, R. J. Cava, *Chem. Mater.* **2023**, *35*, 6005.
[46] M. Ghasemi, P. Karsili, A. Mishra, M. R. Golobostanfard, J. V. Milić, *Advanced Energy Materials* **2025**, 2502693.
[47] M. Lu, X. Shen, B. Ke, T. Huang, O. Xu, L. Kong, X. Zhong, B. Zou, *Mater. Today Chem.* **2024**, *38*, 102043.
[48] M. Morita, M. Kameyama, *J. Lumin.* **1981**, *24–25*, 79.
[49] K. Zhang, E. Kang, R. Huang, L. Li, Y. Wang, H. Zhao, M. Hagiwara, Y. Ma, Y. Han, *Adv. Opt. Mater.* **2024**, 2400936.
[50] X. Li, J. M. Hoffman, M. G. Kanatzidis, *Chem. Rev.* **2021**, *121*, 2230.
[51] M. Seitz, A. J. Magdaleno, N. Alcázar-Cano, M. Meléndez, T. J. Lubbers, S. W. Walraven, S. Pakdel, E. Prada, R. Delgado-Buscalioni, F. Prins, *Nat. Commun.* **2020**, *11*, 2035.
[52] M. E. Kamminga, R. Hidayat, J. Baas, G. R. Blake, T. T. M. Palstra, *APL Mater.* **2018**, *6*, 066106.
[53] Tsuchiya, Naoto, Aoki, Saya, Nakayama, Yuki, Cosquer, Goulven, Nishihara, Sadafumi, Pardo-Sainz, Miguel, Rodríguez-Velamazán, José Alberto, Campo, Javier, Inoue, Katsuya, *CCDC 2411272: Experimental Crystal Structure Determination*, Cambridge Crystallographic Data Centre.
[54] W. Depmeier, *Acta Crystallogr. B* **1976**, *32*, 303.
[55] W. Depmeier, *Acta Crystallogr., Sect. B: Struct. Sci.* **1977**, *33*, 3713.
[56] S.-H. Park, I.-H. Oh, S. Park, Y. Park, J. H. Kim, Y.-D. Huh, *Dalton Trans.* **2012**, *41*, 1237.
[57] L. Septiany, D. Tulip, M. Chislov, J. Baas, G. R. Blake, *Inorg. Chem.* **2021**, *60*, 15151.
[58] B. Huang, J.-Y. Zhang, R.-K. Huang, M.-K. Chen, W. Xue, W.-X. Zhang, M.-H. Zeng, X.-M. Chen, *Chem. Sci.* **2018**, *9*, 7413.
[59] G. Racah, *Phys. Rev.* **1942**, *61*, 186.
[60] G. Racah, *Phys. Rev.* **1942**, *62*, 438.

[61] A. S. Marfunin, N. G. Egorova, A. G. Mishchenko, *Physics of minerals and inorganic materials: an introduction*, Springer-Verlag, Berlin Heidelberg New York, **1979**.
[62] E. U. Condon, G. Shortley, *The theory of atomic spectra*, Cambridge University press, Cambridge New York Melbourne, **1991**.
[63] B. N. Figgis, *Introduction to ligand fields*, R. E. Krieger, Malabar (Fla.), **1986**.
[64] D. B. Mitzi, In *Progress in Inorganic Chemistry* (Ed.: Karlin, K. D.), Wiley, **1999**, pp. 1–121.
[65] X. Gong, O. Voznyy, A. Jain, W. Liu, R. Sabatini, Z. Piontkowski, G. Walters, G. Bappi, S. Nokhrin, O. Bushuyev, M. Yuan, R. Comin, D. McCamant, S. O. Kelley, E. H. Sargent, *Nat. Mater.* **2018**, *17*, 550.
[66] K. Robinson, G. V. Gibbs, P. H. Ribbe, *Science* **1971**, *172*, 567.
[67] K. Momma, F. Izumi, *Journal of Applied Crystallography* **2011**, *44*, 1272.
[68] W. D. Van Amstel, L. J. De Jongh, *Solid State Commun.* **1972**, *11*, 1423.
[69] Z. Chen, J. Xue, Z. Wang, H. Lu, *Mater. Chem. Front.* **2024**, *8*, 210.
[70] M. P. Davydova, L. Meng, M. I. Rakhmanova, I. Yu. Bagryanskaya, V. S. Sulyaeva, H. Meng, A. V. Artem'ev, *Adv. Opt. Mater.* **2023**, *11*, 2202811.
[71] H. Bahmani Jalali, A. Pianetti, J. Zito, M. Imran, M. Campolucci, Y. P. Ivanov, F. Locardi, I. Infante, G. Divitini, S. Brovelli, L. Manna, F. Di Stasio, *ACS Energy Lett.* **2022**, *7*, 1850.
[72] S. Sugano, *Multiplets of Transition-Metal Ions in Crystals*, Elsevier Science, Saint Louis, **2014**.
[73] K. Xu, A. Meijerink, *Chem. Mater.* **2018**, *30*, 5346.
[74] S. Ithurria, P. Guyot-Sionnest, B. Mahler, B. Dubertret, *Phys. Rev. Lett.* **2007**, *99*, 265501.
[75] M. C. Day, J. Selbin, *Theoretical inorganic chemistry*, Reinhold Book Corp., **1927**.
[76] Y. Tanabe, S. Sugano, *J. Phys. Soc. Jpn.* **1954**, *9*, 753.
[77] Y. Tanabe, S. Sugano, *J. Phys. Soc. Jpn.* **1954**, *9*, 766.
[78] D. Castañeda, G. Muñoz H., U. Caldiño, *Opt. Mater.* **2005**, *27*, 1456.
[79] G. K. B. Costa, S. S. Pedro, I. C. S. Carvalho, L. P. Sosman, *Opt. Mater.* **2009**, *31*, 1620.
[80] D. Curie, C. Barthou, B. Canny, *J. Chem. Phys.* **1974**, *61*, 3048.
[81] R. E. Trees, *Phys. Rev.* **1951**, *83*, 756.
[82] A. K. Mehra, *J. Chem. Phys.* **1968**, *48*, 4384.
[83] F. E. Mabbs, D. J. Machin, *Magnetism and transition metal complexes*, Dover ed., Dover Publications, Mineola, N.Y, **2008**.
[84] J. F. Suyver, J. J. Kelly, A. Meijerink, *J. Lumin.* **2003**, *104*, 187.
[85] G. Mackh, W. Ossau, D. R. Yakovlev, A. Waag, G. Landwehr, R. Hellmann, E. O. Göbel, *Phys. Rev. B* **1994**, *49*, 10248.
[86] H. D. Nelson, L. R. Bradshaw, C. J. Barrows, V. A. Vlaskin, D. R. Gamelin, *ACS Nano* **2015**, *9*, 11177.
[87] F. Godejohann, R. R. Akhmadullin, K. V. Kavokin, D. R. Yakovlev, I. A. Akimov, B. R. Namozov, Yu. G. Kusrayev, G. Karczewski, T. Wojtowicz, M. Bayer, *Phys. Rev. B* **2022**, *106*, 195305.
[88] J. J. Schuyt, G. V. M. Williams, S. V. Chong, *Opt. Mater.: X* **2024**, *23*, 100345.

**Supporting Information**

Photographs and XRD of the synthesized crystals. Average atomic displacements. Temperature-dependent PLE spectra in the 7-270 K range. Tanabe-Sugano diagram for $Mn^{2+}$ compounds. Temperature-dependent C Racah parameter and PL spectra in the 100-300 K range.

Background from PLE sample holder. PLE data analysis. Supporting Information is available from the Wiley Online Library or from the author.